\documentclass{caosp305}

\usepackage{graphicx}
\usepackage{natbib}
\articleNo{123}
\pubyear{2018}
\volume{35}
\volnumber{3}
\firstpage{1}
\received{November 10, 2017}
\accepted{November 10, 2017}

\def\BibTeX{{\rm B\kern-.05em{\sc i\kern-.025em b}\kern-.08em
             T\kern-.1667em\lower.7ex\hbox{E}\kern-.125emX}}

\begin{document}

\htitle{Interacting Binaries}
\hauthor{G.\,Briggs, {\it et al.}}

\title{Magnetic Fields in Interacting Binaries}

\author{
        G.\,Briggs \inst{1,}
        \and
        Lilia Ferrario \inst{1,}
        \and
       C.\,A.\,Tout \inst{2,} 
       \and 
       D.\,T.\,Wickramasinghe \inst{1,}
       }

\institute{
           Mathematical Sciences Institute, The Australian National
           University, \\
            Australia, \email{Gordon.Briggs@anu.edu.au}\\
           \and
           Institute of Astronomy, University of Cambridge, U.K.\\
          }

\date{November 10, 2017}

\maketitle

\begin{abstract}
  \citet{WTF2014} and \citet{Briggs2015} have proposed that the strong
  magnetic fields observed in some single white dwarfs (MWDs) are
  formed by an $\alpha$--$\Omega$ dynamo driven by differential rotation
  when two stars, the more massive one with a degenerate core, merge
  during common envelope (CE) evolution \citep{Ferrario2015Origin}. We
  synthesise a population of binaries to investigate if fields in the
  magnetic cataclysmic variables (MCVs) may also originate during
  stellar interaction in the CE phase.

  \keywords{magnetic field -- white dwarfs -- binaries -- cataclysmic variables}

\end{abstract}

\section{Methods and Discussion} 

In the MCVs, a red dwarf transfers matter to a MWD via magnetically
confined accretion flows.  Cyclotron and Zeeman spectroscopy have
revealed fields of a few $10^7$--$10^8$\,G
\citep[][]{Ferrario1992, Ferrario1993a, Ferrario1996} in the high
field MCVs \citep[the polars;][]{Ferrario1999}.  Fields of a few
$10^6$--$10^7$\,G are inferred in the intermediate polars
\citep[][]{Ferrario1993b,Ferrario1993c}.

We have synthesised a population of binaries with the BSE code of
\citet{Hurley2002} for a CE efficiency parameter $\alpha$ in the range
0.1--0.9. We have assumed that the field $B$, achieved by the WD
during CE evolution, is proportional to the orbital angular
velocity $\Omega$ of the binary when the envelope gets ejected. If
$10^{13}$\,G is the highest field that can be stably generated in a WD
then
\begin{equation}\label{EqBfield}
B = \gamma 10^{13}~\left(\frac{\Omega}{\Omega_{\rm crit}}\right)\, \mbox{G}.
\end{equation}
where $\Omega_{\rm crit}$ is the break-up angular velocity of the WD
and $\gamma$ is a parameter that determines the efficiency with which
the poloidal field is regenerated by the decaying toroidal field. The
best fit to observations requires $\gamma\sim 10^{-3}$.
\begin{figure}
\begin{minipage}[r]{0.5\linewidth}
\includegraphics[width=\linewidth]{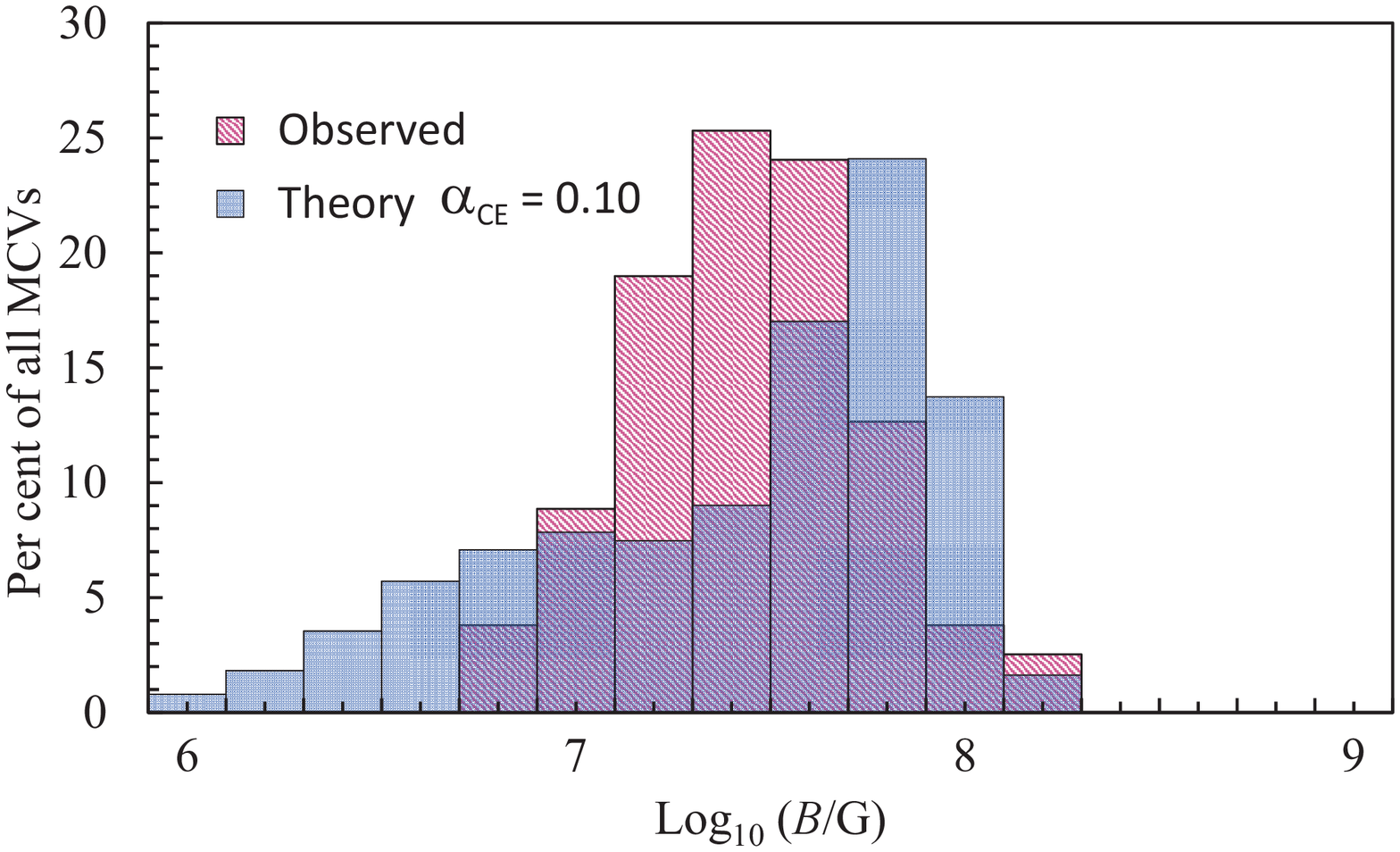}
\end{minipage}
\hfill
\begin{minipage}[l]{0.5\linewidth}
\includegraphics[width=\linewidth]{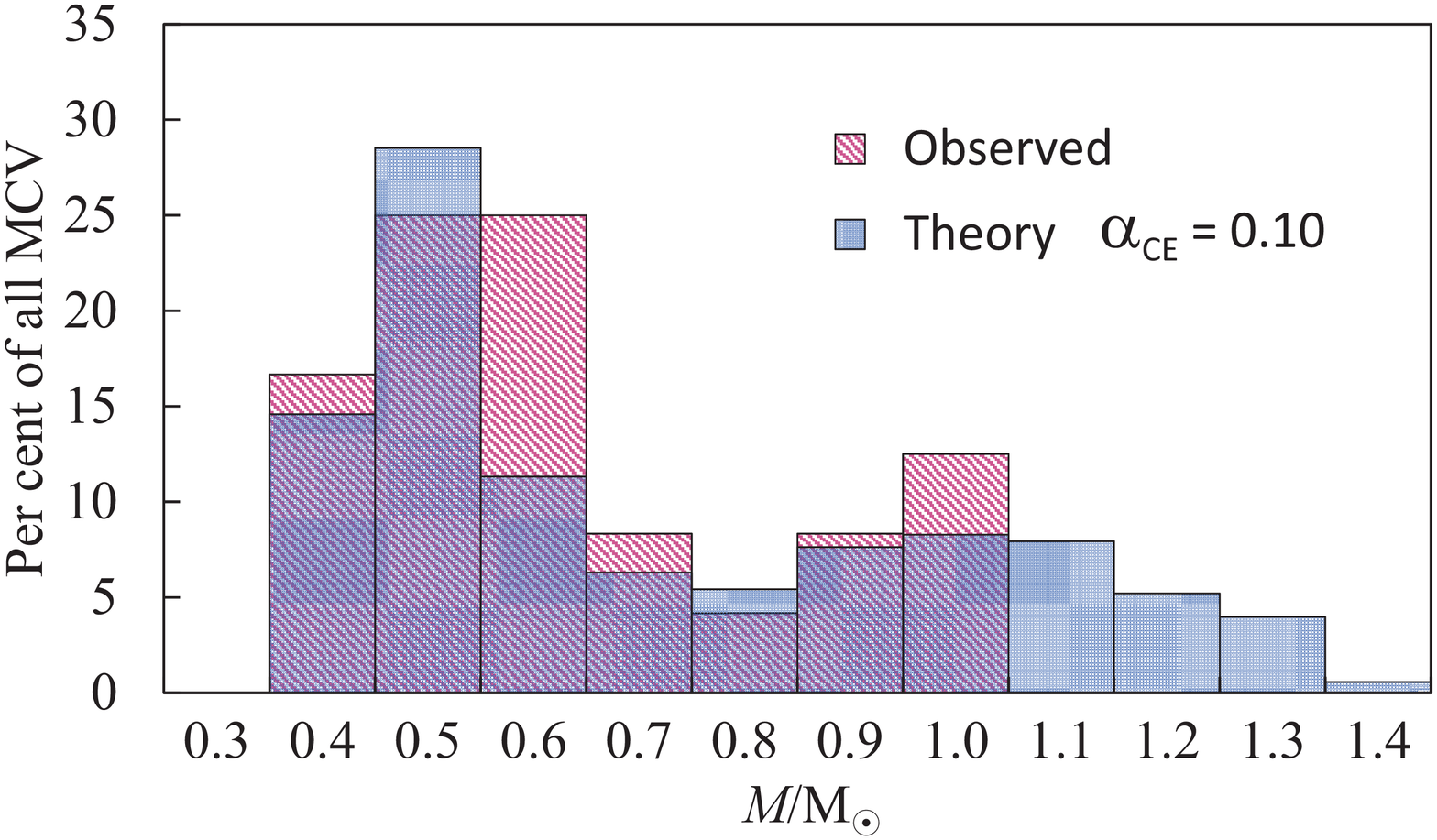}
\end{minipage}%
\caption{Left: Comparison of the theoretical field strength for
  $\alpha= 0.1$ and observations from \citet{Ferrario2015MWD}. Right:
  Comparison of the theoretical mass distributions to observations
  from \citet{Zorotovic2011}.}
\label{comparison}
\end{figure}

We find that if $\alpha<0.4$ we can produce binaries emerging from the CE
that are close to contact, in agreement with \citet{Schwope2009} who
proposed that those detached magnetic binaries where the MWD accretes
matter from the wind of its companion are the progenitors of the
MCVs. The theoretical and observed field and mass distributions are
overlapped and shown in Fig.\,\ref{comparison}.

\acknowledgements L.F. acknowledges support from the Grant Agency of
the Czech Republic (15-15943S) and the organisers of the conference.

\bibliography{Briggs_MCVs}

\begin{thebibliography}{13}
\expandafter\ifx\csname natexlab\endcsname\relax\def\natexlab#1{#1}\fi

\bibitem[{{Briggs} {et~al.}(2015){Briggs}, {Ferrario}, {Tout},
  {Wickramasinghe}, \& {Hurley}}]{Briggs2015}
{Briggs}, G.~P., {Ferrario}, L., {Tout}, C.~A., {Wickramasinghe}, D.~T., \&
  {Hurley}, J.~R. 2015, {\it \mnras}, {\bf 447}, 1713, DOI
  10.1093/mnras/stu2539

\bibitem[{{Ferrario} {et~al.}(1996){Ferrario}, {Bailey}, \&
  {Wickramasinghe}}]{Ferrario1996}
{Ferrario}, L., {Bailey}, J., \& {Wickramasinghe}, D. 1996, {\it \mnras}, {\bf
  282}, 218, DOI 10.1093/mnras/282.1.218

\bibitem[{{Ferrario} {et~al.}(1993{\natexlab{a}}){Ferrario}, {Bailey}, \&
  {Wickramasinghe}}]{Ferrario1993a}
{Ferrario}, L., {Bailey}, J., \& {Wickramasinghe}, D.~T. 1993{\natexlab{a}},
  {\it \mnras}, {\bf 262}, 285, DOI 10.1093/mnras/262.2.285

\bibitem[{{Ferrario} {et~al.}(2015{\natexlab{a}}){Ferrario}, {de Martino}, \&
  {G{\"a}nsicke}}]{Ferrario2015MWD}
{Ferrario}, L., {de Martino}, D., \& {G{\"a}nsicke}, B.~T. 2015{\natexlab{a}},
  {\it \ssr}, {\bf 191}, 111, DOI 10.1007/s11214-015-0152-0

\bibitem[{{Ferrario} {et~al.}(2015{\natexlab{b}}){Ferrario}, {Melatos}, \&
  {Zrake}}]{Ferrario2015Origin}
{Ferrario}, L., {Melatos}, A., \& {Zrake}, J. 2015{\natexlab{b}}, {\it \ssr},
  {\bf 191}, 77, DOI 10.1007/s11214-015-0138-y

\bibitem[{{Ferrario} \& {Wehrse}(1999)}]{Ferrario1999}
{Ferrario}, L. \& {Wehrse}, R. 1999, {\it \mnras}, {\bf 310}, 189, DOI
  10.1046/j.1365-8711.1999.02976.x

\bibitem[{{Ferrario} \& {Wickramasinghe}(1993)}]{Ferrario1993b}
{Ferrario}, L. \& {Wickramasinghe}, D.~T. 1993, {\it \mnras}, {\bf 265}, 605,
  DOI 10.1093/mnras/265.3.605

\bibitem[{{Ferrario} {et~al.}(1992){Ferrario}, {Wickramasinghe}, {Bailey},
  {Hough}, \& {Tuohy}}]{Ferrario1992}
{Ferrario}, L., {Wickramasinghe}, D.~T., {Bailey}, J., {Hough}, J.~H., \&
  {Tuohy}, I.~R. 1992, {\it \mnras}, {\bf 256}, 252, DOI
  10.1093/mnras/256.2.252

\bibitem[{{Ferrario} {et~al.}(1993{\natexlab{b}}){Ferrario}, {Wickramasinghe},
  \& {King}}]{Ferrario1993c}
{Ferrario}, L., {Wickramasinghe}, D.~T., \& {King}, A.~R. 1993{\natexlab{b}},
  {\it \mnras}, {\bf 260}, 149, DOI 10.1093/mnras/260.1.149

\bibitem[{{Hurley} {et~al.}(2002){Hurley}, {Tout}, \& {Pols}}]{Hurley2002}
{Hurley}, J.~R., {Tout}, C.~A., \& {Pols}, O.~R. 2002, {\it \mnras}, {\bf 329},
  897, DOI 10.1046/j.1365-8711.2002.05038.x

\bibitem[{{Schwope} {et~al.}(2009){Schwope}, {Nebot Gomez-Moran}, {Schreiber},
  \& {G{\"a}nsicke}}]{Schwope2009}
{Schwope}, A.~D., {Nebot Gomez-Moran}, A., {Schreiber}, M.~R., \&
  {G{\"a}nsicke}, B.~T. 2009, {\it \aap}, {\bf 500}, 867, DOI
  10.1051/0004-6361/200911699

\bibitem[{{Wickramasinghe} {et~al.}(2014){Wickramasinghe}, {Tout}, \&
  {Ferrario}}]{WTF2014}
{Wickramasinghe}, D.~T., {Tout}, C.~A., \& {Ferrario}, L. 2014, {\it \mnras},
  {\bf 437}, 675, DOI 10.1093/mnras/stt1910

\bibitem[{{Zorotovic} {et~al.}(2011){Zorotovic}, {Schreiber}, \&
  {G{\"a}nsicke}}]{Zorotovic2011}
{Zorotovic}, M., {Schreiber}, M.~R., \& {G{\"a}nsicke}, B.~T. 2011, {\it \aap},
  {\bf 536}, A42, DOI 10.1051/0004-6361/201116626

\end{thebibliography}

\end{document}